\documentclass{article}
\usepackage{graphicx} 
\usepackage[a4paper, total={6in, 8in}]{geometry}
\usepackage{hyperref}
\usepackage[numbers,sort&compress]{natbib}
\usepackage{amsmath}
\usepackage{amssymb}

\title{\Large{\textbf{Weak Bending of Light by Rotating Regular Black Holes with Asymptotically Minkowski Core using the Gauss-Bonnet Theorem}}}

\author{Miles Angelo P. Sodejana}
\author{{Miles Angelo P. Sodejana } \\ {Department of Physics, University of Southern Mindanao } \\ {9407 Kabacan, Cotabato, Philippines} \\ {Email: mapsodejana@usm.edu.ph}}
\date{October 2024}

\begin{document}

\maketitle

\begin{abstract}
In this paper, the weak gravitational lensing phenomenon for a recently proposed rotating regular black hole with an asymptotically Minkowski core characterized by a sub-Planckian curvature was investigated.  Using the Gauss-Bonnet Theorem, the deflection of light in the weak limit was computed by taking the black hole as a lens at a finite distance from both the source and the observer.  It was shown that the weak deflection angle slightly differs between the prograde and retrograde motion but both eventually converge to $0$ as $b$ increases. Moreover, the deflection angle correction for Kerr classical black hole and this sort of rotating regular black hole is a decreasing function for large values of $b$.  It was also shown that the weak deflection angle for this sort of regular black hole is similar to Bardeen and Hayward black hole given its corresponding values for the parameters $x$ and $n$.
\end{abstract}

\section{Introduction}
\hspace*{3mm}The advent of the first black hole shadow images by the Event Horizon Telescope (ETH) collaboration has ushered in a transformative era in black hole astrophysics \cite{EventHorizonTelescope:2019dse, EventHorizonTelescope:2022wkp}.  Despite these advances that further solidified general relativity (GR), spacetime singularities still present a major challenge within black holes and at the universe's inception. A prominent issue in GR is the singularity problem, where the scalar curvature becomes infinite at a black hole's center \cite{Hawking:1966sx, Penrose:1964wq, Goswami:2005fu, Janis:1968zz, Joshi:2011rlc}. It is theoretically believed that the divergence of the Kretschmann scalar curvature signals the breakdown of classical general relativity in extreme environments. In such cases, the singularity at the center of a classical black hole could potentially be resolved or avoided through the influence of quantum gravitational effects \cite{Ali:2015tva, Calmet:2017qqa, Callan:1992rs, DeWitt:1967ub, DeWitt:1967uc, DeWitt:1967yk, Donoghue:1993eb, tHooft:1984kcu, Garay:1994en, Han:2004wt}.  While quantum corrections are believed to potentially resolve these singularities, a comprehensive theory of quantum gravity remains elusive. \\
\hspace*{3mm} Historically, resolving this singularity problem has been dealt with leading to the proposal of regular black holes.  Literature has presented phenomenological models of such regular black holes, which can be categorized based on their behavior near their cores. One category includes regular black holes with an asymptotically de Sitter core, such as the Bardeen, Hayward, and Frolov black holes \cite{1968qtr..conf...87B, Frolov:2014jva, Hayward:2005gi}. Another category features black holes with an asymptotically Minkowskian core, distinguished by an exponentially suppressed Newtonian potential \cite{Culetu:2013fsa, Li:2016yfd, Ling:2021olm, Simpson:2019mud, Martinis:2010zk, Xiang:2013sza, Ben-Amots:2011lbe}. Recently, a new class of regular black holes with an asymptotically Minkowskian core was proposed \cite{Ling:2021olm}. These black holes have scalar curvatures that remain finite and sub-Planckian throughout their evaporation process, regardless of their mass. This behavior is consistent with the expectations of quantum gravity, which posits that the energy scale of objects should be constrained by the Planck energy within quantum gravitational theory.  The correspondence between the regular black holes with asymptotically dS core and those with asymptotically Minkowski core was also discussed in \cite{Ling:2021olm}, and most importantly, the rotating case was discussed in \cite{Ling:2022vrv}.  \\
\hspace*{3mm} The optical properties of black holes have been crucial in understanding phenomena in strong gravity regimes, mainly focusing on their shadow and deflection angle. In 1919, the first evidence for Einstein's general relativity manifested through the observation of gravitational deflection of light by the Sun \cite{Dyson:1920cwa}, and ever since then, gravitational lensing, a phenomenon defined as the bending of light by the presence of matter and energy, has been thoroughly investigated in both cosmology and astronomy \cite{Bartelmann:2010fz, Blandford:1991xc, Wambsganss:1998gg}.  Moreover, gravitational lensing in strong and weak limits \cite{Geiller:2020xze, Brahma:2020eos, Liu:2020ola} have also been vital as astrophysical tools \cite{Aubourg:1993wb} in investigating strong gravitational fields and dark matter detection \cite{Clowe:2006eq}.  In this lensing phenomenon, the deflection angle of light is influenced by the physical properties of the lensing object (such as a black hole) and the distance between the observer and the lens. This phenomenon allows researchers to study the characteristics of black holes through their lensing effects, offering a method to differentiate between various types of black holes \cite{Wang:2019cuf, Bin-Nun:2009hct, Zhao:2016kft, Zhao:2017cwk, Wei:2011nj, Chakraborty:2016lxo, Eiroa:2002mk, Kraniotis:2010gx, Kraniotis:2014paa, Kumar:2020sag, Jusufi:2018jof, Jusufi:2019caq, Ghosh:2020spb, Gyulchev:2006zg} and test and constrain different theories of gravity \cite{Allahyari:2019jqz, Badia:2017art, Sahu:2015dea, Kuang:2022ojj, Kuang:2022xjp, Li:2020zxi, Afrin:2021imp, Afrin:2021wlj, Bhadra:2003zs, Horvath:2011xr, Pantig:2022ely, Soares:2023err, Vagnozzi:2022moj}. \\
\hspace*{3mm} A large number of photons passing around black holes reveal a dark shadow, a photon sphere, and relativistic images due to the gravitational lensing effect at the horizon. This phenomenon of a black hole shadow through lensing led to a surge in related research on the topic \cite{Cunha:2018acu, Cunha:2015yba, Younsi:2016azx, Ghosh:2022gka, Konoplya:2019fpy, Konoplya:2021slg, Vagnozzi:2019apd, Gralla:2019xty, Shaikh:2018kfv, Shaikh:2018lcc, Tsukamoto:2017fxq, Abdujabbarov:2015pqp, Abdujabbarov:2016hnw, Bambi:2008jg, Grenzebach:2014fha, Amarilla:2011fx, Hioki:2009na, Takahashi:2004xh, Falcke:1999pj}.   When photons are located far from the black hole, gravitational lensing is best described using the weak field limit. Pioneering this approach, Gibbons and Werner applied the Gauss-Bonnet theorem (GBT) to the optical metric of a spherically symmetric black hole, calculating the light deflection angle in the weak field regime \cite{Gibbons:2008rj}. Subsequently, Werner extended this analysis by employing the osculating Riemann approach within the framework of Finsler geometry to explore lensing effects in Kerr black holes \cite{Werner:2012rc}. However, Finsler geometry is not well-suited for calculating finite-distance corrections. To address this limitation, Ono and collaborators analyzed the weak deflection angle of Kerr black holes in spatial geometry, specifically to test finite-distance corrections in axially symmetric spacetimes \cite{Ono:2017pie}, generalizing the work Ishihara et al in extending the calculations for the weak deflection angle using GBT for finite distance for a static, spherically symmetric, and asymptotically flat spacetime \cite{Ishihara:2016vdc}.  Furthermore, this method has been used to discuss the weak deflection angle by regular black holes with dS core and its modifications \cite{Jusufi:2018jof, Belhaj:2022vte, Ovgun:2019wej}. \\ 
\hspace*{3mm}
This paper investigates how the deviation parameter $\alpha_0$ and the spin parameter $a$ influence the weak field limits of gravitational lensing by rotating regular black holes with asymptotically Minkowski core and sub-Planckian curvature by Ling and Wu \cite{Ling:2022vrv}, employing the Gauss-Bonnet theorem (GBT) extended by Ono et. al \cite{Ono:2017pie}. The structure of the paper is organized as follows: Section 2 provides a brief overview of the rotating regular black hole with an asymptotically Minkowski core and sub-Planckian curvature and its null geodesic. Section 3 reviews the Gauss-Bonnet theorem and uses it to analyze the light deflection in the weak field limit for our black hole metric of interest.  The paper concludes with a summary and discussion of the findings in Section 4.
\section{Spacetime and Null Geodesic}
\hspace{3mm} In \cite{Li:2016yfd}, a new sort of regular spherically symmetric black hole was proposed by Ling and Wu with a metric given as  
\begin{align}
    ds^2 = - f(r) dt^2 + f(r)^{-1} dr^2 + r^2 (d\theta^2 + \sin^2 \theta d\phi^2), \label{1}
\end{align}with 
\begin{align}
    f(r) = 1 - \frac{2m(r)}{r}, \label{2}
\end{align} where $m(r)$ is expressed as
\begin{align}
    m(r) = Me^{-\alpha_0 M^x/r^n}. \label{3}
\end{align}This paper uses the geometrized unit G = c = 1.
\hspace{3mm}The metric described can be viewed as a solution to the Einstein field equations coupled to a nonlinear Maxwell field. This indicates that the origin of a regular black hole could be attributed to a nonlinear electromagnetic field \cite{Ayon-Beato:1998hmi}.  A more thorough discussion on the stress-energy tensor and the violation of the strong energy condition is presented in \cite{Li:2016yfd}. \\
\hspace*{3mm} In these sorts of regular black holes, the exponentially suppressing form of the Newton potential leads to a non-singular Minkowski core at the center of the black hole, as originally proposed in \cite{Xiang:2013sza}, but with a specific form of $x = 0$ and $n = 2$.  It was also pointed out in \cite{Li:2016yfd} that its Kretschmann scalar curvature is always sub-Planckian regardless of the mass of the black hole if it satisfies the condition $n \geq x \geq n/3$ and $n \geq 2$ for a fixed $\alpha_0$, which was found in \cite{Zeng:2022yrm} to be $0 \leq \alpha_0 \leq 0.73$.  This condition guarantees the existence of the horizon of the black hole.  A one-to-one correspondence also exists between this sort of regular black holes and the ones with asymptotically de Sitter cores as discussed in \cite{Li:2016yfd}, such that $m(r)$ took the form 
\begin{align}
    m(r) = \frac{Mr^{\frac{n}{x}}}{(r^n + x\alpha_0 M^x)^{1/x}}, \label{4}
\end{align}in which taking $x = 2/3, \ n = 2$ produces a Bardeen black hole, while $x = 1, \ n =3$ produces a Hayward black hole.\\
\hspace*{3mm}Utilizing the Newman-Janis algorithm \cite{Newman:1965tw, Azreg-Ainou:2014aqa, Azreg-Ainou:2014nra, Azreg-Ainou:2014pra, Wu:2022ydk, Toshmatov:2014nya, Ghosh:2014pba}, Ling and Wu generalized the metric in (\ref{1}) to describe a rotating Kerr-like black hole with the following metric \cite{Ling:2022vrv}:
\begin{align}
    ds^2 = -A(r, \theta) dt^2 - 2H(r, \theta)dtd\phi + B(r, \theta)dr^2 + C(r, \theta)d\theta^2 + D(r, \theta)d\phi^2, \label{5}
\end{align} with 
\begin{align} 
A(r, \theta) &= 1 - \frac{2m(r)r}{\Sigma}, \hspace{10mm} H(r, \theta) = \frac{2am(r)r\sin^2 \theta}{\Sigma}, \hspace{10mm} B(r,\theta) = \frac{\Sigma}{\Delta},\notag \\ C(r, \theta) &=\Sigma, \hspace{10mm} D(r, \theta) = \left(r^2 + a^2 + \frac{2a^2 m(r)r\sin^2 \theta}{\Sigma }\right)\sin^2 \theta d\phi^2  \label{6}
\end{align}
where $\Sigma =r^2 + a^2 \cos^2 \theta$ and $\Delta = r^2 - 2m(r)r + a^2$, respectively.  Here, $a$ is the rotation parameter, where it can be seen that (\ref{6}) becomes (\ref{1}) when $a \rightarrow 0$.  It can also be seen that (\ref{1}) and (\ref{6}) become the classical Schwarzchild and Kerr black holes when $\alpha_0 \rightarrow 0$. \\
\hspace*{3mm} The Lagrangian of photons in the metric above is defined as $\mathcal{L} = \dfrac{1}{2}g_{\mu \nu} \dot{x^{\mu}}\dot{x^{\nu}} = 0 $, where the overdot implies a derivative with respect to the affine parameter $\lambda$ for the null geodesic.  We also consider our system to be on the equatorial plane $(\theta = \pi/2).$ Since the metric is stable and axially symmetrical, we have two conserved quantities, i.e., the energy and angular momentum such that 

\begin{align}
     E &= \frac{\partial \mathcal{L}}{\partial \dot{t}} = -g_{tt}\dot{t} - g_{t\phi} \dot{\phi}, \label{7} \\ L &= -\frac{\partial \mathcal{L}}{\partial \dot{\phi}} = g_{t\phi} \dot{t} + g_{\phi \phi} \dot{\phi}. \label{8}
\end{align}
\hspace*{3mm} Setting $E = 1$ and $L/E = b$ where $b$ is the impact parameter, we obtain the following equations of motion for photon trajectory: 
\begin{align}
    \dot{t} & = \frac{D - Hb}{AD + H^2}, \label{9} \\
    \dot{\phi} &= \frac{H + Ab}{AD + H^2}, \label{10} \\
    \dot{r}^2 &= \frac{D - 2Hb - Ab^2}{B(AD + H^2)}. \label{11}
\end{align}

\section{Bending of light in the weak field limit}
\begin{figure}[!t]
    \centering
    \includegraphics[width=0.6\linewidth]{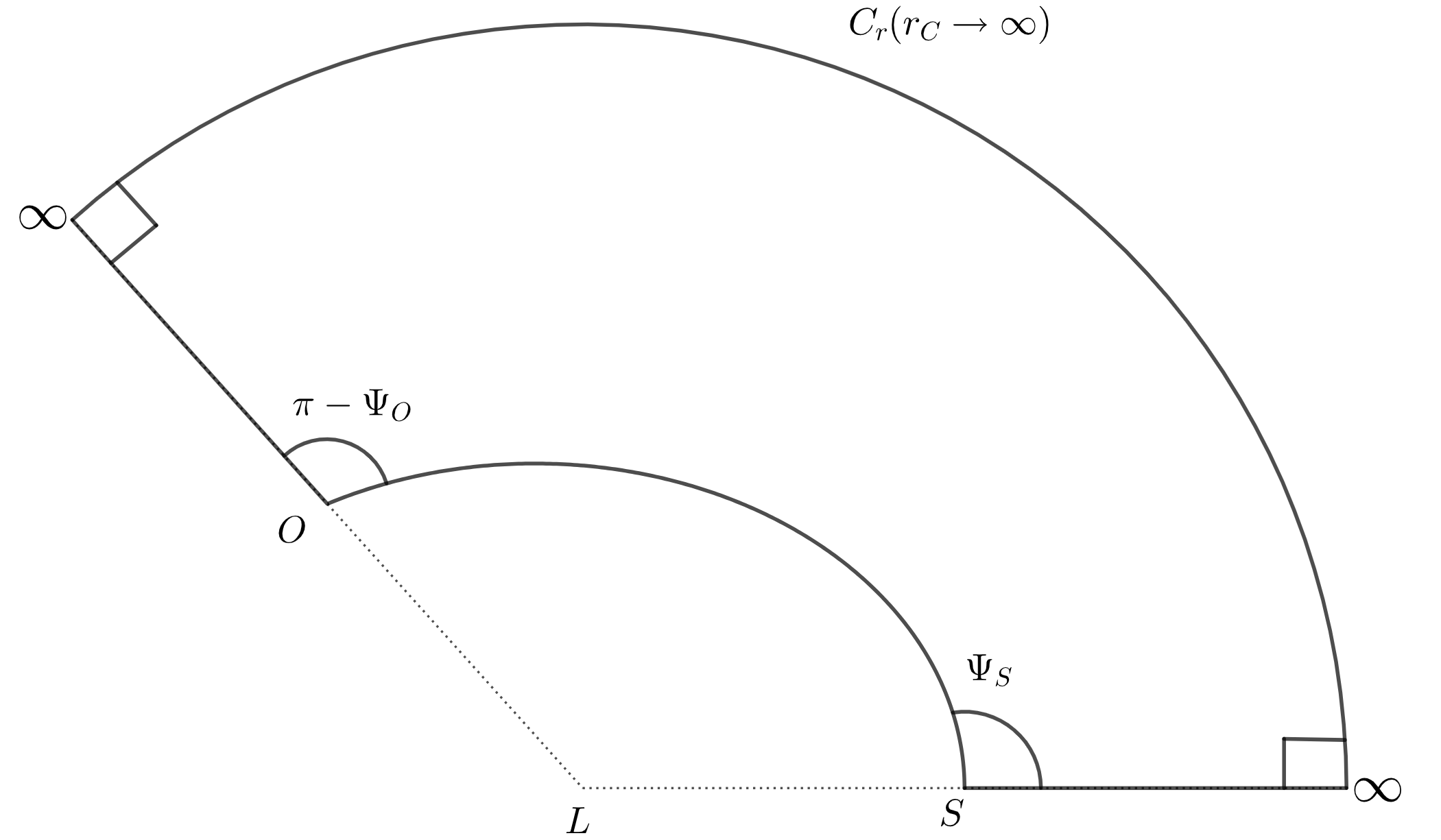}
    \caption{Quadrilateral $^{\infty}_O \Box ^{\infty}_S$ embedded in  a curved space.  The inner angle at the vertex $O$ is $\pi - \Psi_O$.}
    \label{fig: 1}
\end{figure}

\hspace*{3mm}In the weak field limit, we consider situations 
\begin{align}
    \frac{M}{r_0} \ll 1, \hspace{3mm} \frac{M}{b} \ll 1, \label{12}
\end{align}such that the mass is much smaller than the distance scale.  Here, $r_0$ is the closest approach distance from the black hole while $b$ is the impact parameter.  The condition in \eqref{12} tells us that the impact parameter of the light rays is large hence the closest approach to the black hole is much larger than the radius of the light orbit.  This situation says that the deflection angle is much smaller than $2\pi$. In this scenario, we calculate the deflection angle in the weak limit using the Gauss-Bonnet Theorem (GBT) approach.  \\
\hspace*{3mm} Initially, Gibbon and Werner used the GBT to compute the light deflection angle in the weak field limit for spherically symmetric black hole spacetimes \cite{Gibbons:2008rj}. Werner \cite{Werner:2012rc} and Ono et. al \cite{Ono:2017pie} then expanded this approach to Kerr spacetime, applying it to Kerr-Randers optical geometry and spatial metrics. In this section, we use the method extended by Ono et. al \cite{Ono:2017pie}.\\
\hspace*{3mm} A black hole can be modeled as a lens (L) located at a finite distance from both the observer (O) and the source (S).  The deflection angle of light can be calculated along the equatorial plane ($\theta = \pi/2$) by the formula \cite{Ono:2017pie, Ishihara:2016vdc} 
\begin{align}
    \hat{\Theta}= \Psi_O - \Psi_S + \Phi_{OS} \label{13}
\end{align}
where $\Phi_{OS} = \Phi_O - \Phi_S$ is the observer-source angular separation, while $\Psi_S$ and $\Psi_O$ are the angular coordinates at $O$ and $S$. Fig. \ref{fig: 1} gives the visualization of the scenario. The quadrilateral $^{\infty}_{O} \Box ^{\infty}_{S}$ that is embedded in a 3-dimensional manifold $^{(3)} \mathcal{M}$, is consisting of spatial light ray curve from $S$ to $O$, two outgoing radial lines $O$ and from $S$, and a circular arc segment $C_r$ of coordinate radius $r_C$ $(r_C \rightarrow \infty)$.  Applying the GBT to this quadrilateral, the deflection angle \eqref{13} can be written as \cite{Ono:2017pie}
\begin{align}
    \hat{\Theta}=-\int \int_{^{\infty}_{O}\Box ^{\infty}_{S}} \mathcal{K} dS + \int_{S}^{O} k_g dl, \label{14}
\end{align}
where $\mathcal{K}$ and $k_g$ denote the Gaussian curvature of the surface of light propagation and the light curve's geodesic curvature, while $dS$ is the infinitesimal surface area element and $dl$ is the infinitesimal arc line element. When $dl > 0$, the photons are in prograde motion and retrograde when $dl < 0$.  To obtain the weak light deflection angle near the black hole, we first have to solve for the Gaussian curvature $\mathcal{K}$ of the light's path and compute the quadrilateral surface integral of this curvature.  Using Eq. (\ref{5}) for null geodesic $ds^2 = 0$, we derive \cite{Ono:2017pie, Ishihara:2016vdc}
\begin{align}
    dt = \pm\sqrt{\gamma_{ij} dx^i dx^j }+\beta_i dx^i , \label{15}
\end{align}where $\gamma_{ij}$ is the optical metric and $\beta_i$ is the corresponding one-form.  These are expressed as 
\begin{align}
\gamma_{ij} dx^i dx^j &= \frac{B}{A}dr^2+ \frac{C}{A} d\theta^2 + \frac{AD+H^2}{A^2} d\phi^2, \label{16}\\\beta_i dx^i &=-\frac{H}{A} d\phi. \label{17}
\end{align}Using the optical metric, the Gaussian curvature $\mathcal{K}$ of the surface of the light propagation  that relates to the two-dimensional Riemann tensor can be expressed as \cite{Werner:2012rc, Ono:2017pie}
\begin{align}
    \mathcal{K} = \frac{^{(3)}R_{r\phi r\phi}}{\gamma } = \frac{1}{\sqrt{\gamma}}\left[\frac{\partial}{\partial\phi}\left(\frac{\sqrt{\gamma}}{\gamma_{rr}}\ ^{(3)}\Gamma^{\phi}_{rr} \right)-\frac{\partial}{\partial r}\left(\frac{\sqrt{\gamma}}{\gamma_{rr}}\ ^{(3)}\Gamma^{\phi}_{r\phi}  \right)\right], \label{18}
\end{align} where $\gamma = det(\gamma_{ij})$.

\subsection{Weak Lensing for x = 2/3 and n = 2}
\hspace*{3mm}For a rotating regular black hole with an asymptotically Minkowski core (\ref{5}) with $x = 2/3, \ n = 2$, under the weak field limit and slow rotation, the Gaussian curvature $\mathcal{K}$ for the light propagation is found as 
\begin{align}
\mathcal{K} &= -\frac{1}{\sqrt{\gamma}} \frac{\partial}{\partial r}\left[\frac{1}{2\sqrt{\gamma}}\frac{\partial}{\partial r}\left(\gamma_{\phi \phi}\right)\right] \notag \\  &= -\frac{2M}{r^3} + \frac{3M^2}{r^4} +\frac{12\alpha_0 M^{5/3} }{r^5}-\frac{6Ma^2}{r^5}+\mathcal{O}\left(M^{8/3}, \, \alpha_0^2, \, a^3, \, \frac{1}{r^6} \right) . \label{19}
\end{align}
\hspace*{3mm} The surface integral of the Gaussian curvature over the quadrilateral  $^{\infty}_{O} \Box ^{\infty}_{S}$ reads
\begin{align}
    \int \int_{^{\infty}_{O}\Box^{\infty}_{S}} \mathcal{K} dS = \int_{\phi_S}^{\phi_O} \int_{\infty}^{r_0} \mathcal{K} \sqrt{\gamma} dr d\phi, \label{20}
\end{align}where $r_0$ denotes the closest distance to the black hole or the radius of the photon sphere.  To evaluate (\ref{20}), we analyze the photon equations of motion and use (\ref{10}) and (\ref{11}) to obtain the photon orbit equation
\begin{align}
    \left(\frac{dr}{d\phi}\right)^2 = \frac{AD + H^2}{B}\frac{D - 2H b - Ab^2}{(H+ Ab)^2}. \label{21}
\end{align}
\hspace{3mm}We then introduce $u = 1/r$ to reformulate (\ref{21}) as 
\begin{align}
    \left(\frac{du}{d\phi}\right)^2 = \frac{u^4 (AD + H^2)(D-2Hb-Ab^2)}{B(H+Ab)^2}. \label{22}
\end{align}
\hspace{3mm} Under the slow rotation approximation and the weak field limit, we get the photon orbit equation as \cite{Ono:2017pie}
\begin{align}
    u = \frac{\sin\phi}{b} + \frac{M(1+ \cos^2\phi)}{b^2}- \frac{2aM}{b^3}, \label{23}
\end{align}so that Eq. (\ref{20}) becomes 
\begin{align}
     \int \int_{^{\infty}_{O}\Box^{\infty}_{S}} \mathcal{K} dS = - \int_{\phi_S}^{\phi_O} \int_{0}^{u_0} \frac{\mathcal{K}\sqrt{\gamma}}{u^2}  du d\phi. \label{24}
\end{align}In order to simplify our calculations further, we can take 
\begin{align}
    u \approx \frac{\sin\phi}{b}, \label{25}
\end{align}and get 
\begin{align}
\int \int_{^{\infty}_{O}\Box^{\infty}_{S}} \mathcal{K} dS &= -\int_{\phi_S}^{\phi_O} \int_{0}^{\frac{\sin \phi}{b}} \frac{\mathcal{K} \sqrt{\gamma}}{u^2} du d\phi \notag \\  &=-\left(  \frac{2Ma^2 - 4M^{5/3}\alpha_0}{3b^3}-\frac{M^3}{b^3}\right)\left[(1-b^2 u_O^2)^{3/2}+(1-b^2 u_S^2)^{3/2}\right]\notag\\ &- \left(\frac{9M^2 a^2}{16b^3}-\frac{21M^{8/3}\alpha_0}{b^3}\right)\left[u_O \sqrt{1-b^2 u_O^2}\left(1-2u_O^2 b^2\right)+u_S \sqrt{1-b^2 u_S^2}\left(1-2u_O^2 b^2\right)\right] \notag\\ &+\frac{3}{16b^4}\left(9M^2 a^2 + 4M^2 b^2 - 21M^{8/3}\alpha_0 \right)\left[\pi - \arcsin(bu_O)-\arcsin(bu_S)\right] \notag\\ &+\left(\frac{2M}{b}-\frac{3M^3 - 2Ma^2}{b^3}- \frac{4M^{5/3}\alpha_0}{b^3}
\right)\left(\sqrt{1-u_O^2 b^2}+ \sqrt{1-u_S^2 b^2} \right)\notag\\ &+\frac{3}{4b^3}\left(3M^2 a^2 + M^2 b^2-\frac{7M^{8/3}\alpha_0}{2}\right)\left(u_O \sqrt{1-b^2 u_O^2}+ u_S \sqrt{1-b^2 u_S^2}\right)\notag \\ &+ \mathcal{O}\left(\frac{1}{b^5}, \ \alpha_0^2, M^{11/3}, a^3\right),\label{26}
\end{align}where $u_O$ and $u_S$  are the reciprocals of the observer-source distances from the black hole.  Here, the approximation $\cos \phi_O = -\sqrt{1-b^2 u_O^2}$ and $\cos \phi_S = \sqrt{1-b^2 u_S^2}$ are employed.  To determine the geodesic curvature of light, we use the geodesic curvature in the manifold $^{(3)} \mathcal{M}$ expressed as 
\begin{align}
    {k_g} = -\frac{1}{\sqrt{\gamma \gamma^{\theta \theta}}}\beta_{\theta, r}, \label{27}
\end{align}which for our metric (\ref{5}) yields
\begin{align}
k_g=-\frac{2Ma}{r^3}-\frac{2M^2a}{r^4 }+ \frac{6\alpha_0M^{5/3}a}{r^5}-\frac{3aM^3}{r^5} + \mathcal{O}\left( \frac{aM^{8/3}\alpha_0}{r^6}\right). \label{28}
\end{align}
\hspace{3mm}We solve the path integral of the geodesic curvature using a linear approximation of the photon orbit as $r = b/\cos \vartheta$ and $l = b \tan \vartheta$ \cite{Ono:2017pie}.  We therefore compute the geodesic curvature path integral $k_g$ as 
\begin{align}
    \int_{S}^{O} {k_g} \, dl & = \int_{\phi_S}^{\phi_O} \left(-\frac{2Ma}{b^2}\cos \vartheta-\frac{2M^2a}{b^3 }\cos^2 \vartheta+ \frac{6\alpha_0M^{5/3}a}{b^4}\cos^3 \vartheta-\frac{3aM^3}{b^4} \cos^{3} \vartheta\right) d\vartheta \notag \\ &=\left(\frac{3M^3a}{b^4}+\frac{2aM}{b^2}-\frac{6aM^{5/3}\alpha_0}{b^4}\right)\left(\sqrt{1-b^2 u_O^2 }+ \sqrt{1-b^2 u_S^2 }\right) \notag \\ &+\left(\frac{2aM^{5/3}\alpha_0-M^3a}{b^4}\right)\left[\left(1-b^2 u_O^2\right)^{3/2}+ \left(1-b^2 u_S^2\right)^{3/2}\right] \notag \\ &+ \frac{aM^2 }{b^3}\left[\arcsin\left(\sqrt{1-b^2 u_O^2}\right)+\arcsin\left(\sqrt{1-b^2 u_O^2}\right)\right] \notag \\&+\frac{aM^2}{b^3}\left(b u_O \sqrt{1-b^2 u_O^2 }+ b u_S \sqrt{1-b^2 u_S^2 }\right), \label{29}
\end{align}where we used $\sin \phi_O = - \sqrt{1-b^2 u_O^2}$  and $\sin \phi_S =  \sqrt{1-b^2 u_S^2} \,.$  Adding (\ref{26}) and (\ref{29}) to get (\ref{14}) and taking $u_O \rightarrow 0$ and  $u_S \rightarrow 0$ in the distant limit, we get 
\begin{align}
    \hat{\Theta}  \approx \frac{4M}{b} -\frac{16M^{5/3}\alpha_0}{3b^3}\pm \left(\frac{4aM}{b^2}-\frac{8aM^{5/3}\alpha_0}{b^4}\right), \label{30}
\end{align}where the positive sign implies retrograde photon motion and the negative sign implies prograde photon motion. \\
\vspace*{3mm}
\begin{figure*}[!htb]
       \centering
    \begin{minipage}[t]{0.49 \textwidth}
        \centering
        \includegraphics[width = 1\textwidth]{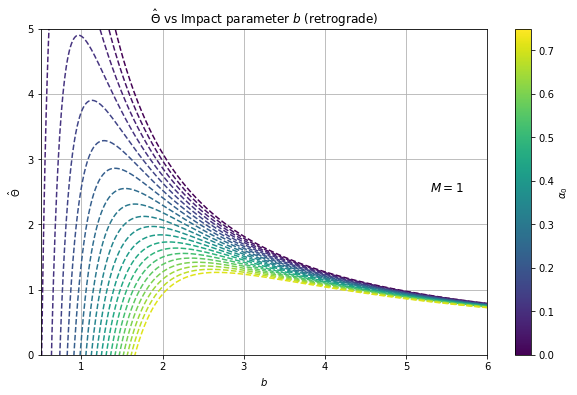}
    \end{minipage}
    \hfill
    \begin{minipage}[t]{0.49 \textwidth}
        \centering
        \includegraphics[width = 1\textwidth]{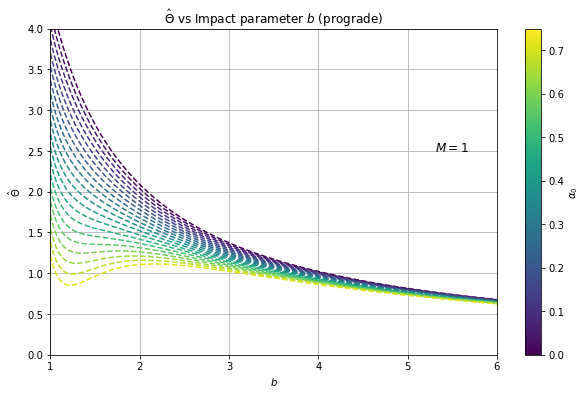}
    \end{minipage}
    \caption{\small{The weak deflection angle $\hat\Theta$ vs. impact parameter $b$ relation for retrograde (left) and prograde (right) motion of photons for the rotating regular black hole with Asymptotically Minkowski core, $x = 2/3, \ n =2$ for $0 \leq \alpha_0  \leq  0.73, \ a = 0.5$ and $M = 1$.}}
    \label{fig: 2}
\end{figure*}
\begin{figure*}[!htb]
       \centering
      \begin{minipage}[t]{0.49 \textwidth}
        \centering
        \includegraphics[width = 1\textwidth]{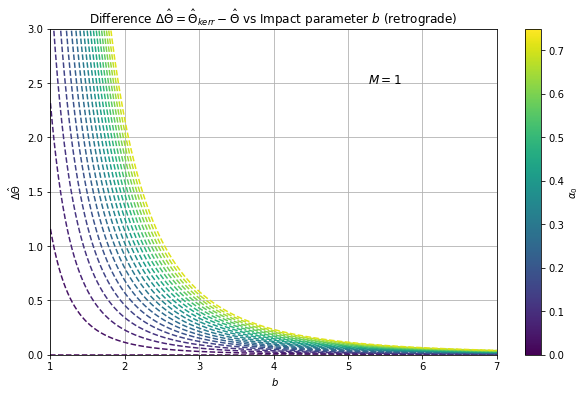}
    \end{minipage}
    \hfill
    \begin{minipage}[t]{0.49 \textwidth}
        \centering
        \includegraphics[width = 1\textwidth]{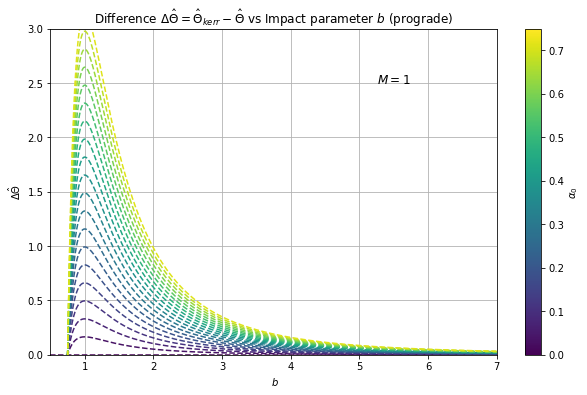}
    \end{minipage}
    \caption{\small{Deflection angle corrections $\delta{\hat\Theta}=\hat\Theta |_{Kerr} - \hat\Theta$ for the weak lensing around the rotating regular black hole with Asymptotically Minkowski core for retrograde (left) and prograde (right) motion of photons, $x = 2/3, \ n =2$ for $0\leq \alpha_0 \leq \ 0.73, \ a = 0.5$ and $M = 1$.}  }
    \label{fig: 3}
\end{figure*}
\hspace*{3mm}In \cite{1968qtr..conf...87B}, Bardeen defined a regular black hole with an asymptotically de Sitter core that can be generalized to a rotating Kerr-like one as (\ref{5}) but with 
\begin{align}
    m(r) = \frac{r^3}{(r^2 + g^2_{\star})^{3/2}}. \label{31}
\end{align}
\hspace*{3mm}Comparing with (\ref{4}) for $x = 2/3, n = 2$, and using (\ref{30}), we get the deflection angle for the weak field limit by the rotating regular Bardeen black hole as \cite{Jusufi:2018jof}
\begin{align}
    \hat\Theta \approx \frac{4M}{b} - \frac{8g^2_\star M}{b^3} \pm \frac{4Ma}{b^2}, \label{32}
\end{align}which is just the result for $ x\alpha_0 M^x = g^2_\star$ for the rotating regular black hole with asymptotically Minkowski core where $x = 2/3, n = 2$.  It shows the correspondence between this sort of regular black hole and that of Bardeen black hole at the weak field limit as discussed in \cite{Li:2016yfd}.  In the limit as $\alpha_0 \rightarrow 0$ and $a \rightarrow 0$, we get the weak deflection angle for the Schwarzschild solution.\\
\hspace*{3mm}In Fig. \ref{fig: 2}, we see that for small values of $b$, the deflection angle is an increasing function with clear variation from different $\alpha_0$ for retrograde motion, while it is a decreasing function for prograde motion.  For large impact parameter values, however, we see that the deflection angle is a decreasing function for both cases, in which the curves converge for varying $\alpha_0$.  In Fig. \ref{fig: 3}, we note that the deflection angle correction for retrograde motion consistently decreases while it initially increases and then decreases for the prograde motion. Eventually, however, the trend converges to $0$ as the impact parameter $b$ increases for both cases. This is expected for larger $r_0$ or farther closest distance of a light ray from the black hole. However, Fig. \ref{fig: 3} can only accurately describe a scenario when $b \gg b_c$, where $b_c$ is the critical impact parameter threshold below which black holes capture the light rays passing around it. When $b > b_c$, the light rays are deflected.

\subsection{Weak Lensing for x = 1 and n = 3}
\hspace{3mm} For a case for the metric (\ref{5}) where $x = 1, \ n = 3$, under the slow rotation approximation and the weak field limit scenario, the Gaussian curvature $\mathcal{K}$ is found as
\begin{align}
    \mathcal{K} &= - \frac{1}{\gamma} \frac{\partial}{\partial r} \left[\frac{1}{2\sqrt{\gamma}} \frac{\partial}{\partial r} (\gamma_{\phi \phi})\right] \notag \\ &= -\frac{2M}{r^3} + \frac{3M^2}{r^4} - \frac{6Ma^2}{r^5} + \frac{12M}{r^5} + \frac{20M^2 \alpha_0 }{r^6} + \mathcal{O}\left(M^3, \alpha_0^2, a^3, \frac{1}{r^7}\right). \label{33}
\end{align}
\hspace{3mm}Following our calculations from Eqs. (\ref{20}) - (\ref{25}), we get 
\begin{align}
    \int \int_{^\infty_O \Box ^\infty_S} \mathcal{K} dS &= -\int_{\phi_S}^{\phi_O} \int_{0}^{\frac{\sin \phi}{b}} \frac{\mathcal{K}\sqrt{\gamma}}{u^2} du d\phi \notag \\ &= -\left(\frac{2Ma^2 - 4M}{b^3} - \frac{M^3}{b^3}\right)\left[(1-b^2 u_O^2)^{3/2} + (1-b^2 u_S^2)^{3/2}\right] \notag \\ &- \left(\frac{9M^2a^2}{16b^3} - \frac{5M^2 \alpha_0}{8b^3} - \frac{9M^2}{8b^3}\right)\left[u_O \sqrt{1-b^2 u_O^2} (1-2u_O^2b^2)+u_S \sqrt{1-b^2 u_S^2} (1-2u_S^2b^2)\right] \notag \\ &+ \left(\frac{27M^2 a^2}{16b^4}+\frac{3M^2}{4b^2} - \frac{15M^2\alpha_0}{8b^4}\right)[\pi - \arcsin(bu_O) - \arcsin(bu_S)] \notag \\ & + \left(\frac{2M}{b} + \frac{2Ma^2 - 3M^3}{b^3} - \frac{4M}{b^3}\right)\left(\sqrt{1-b^2 u_O^2} + \sqrt{1-b^2 u_S^2}\right) \notag \\ & + \left(\frac{9M^2 a^2 }{8b^3}+ \frac{3M^2}{8b}- \frac{5M^2 \alpha_0}{4b^3}-\frac{9M^2}{4b^3}\right)\left(u_0\sqrt{1-b^2 u_O^2} + u_s \sqrt{1-b^2 u_S^2} \right)\notag \\ &+ \mathcal{O}\left(\frac{1}{b^4}, \alpha_0^2, M^4, a^3\right), \label{34}
\end{align}where similar to (\ref{26}), $u_O$ and $u_S$ are the reciprocals of the observer-source distances from the black hole, and we employ the approximation $\cos \phi_O = - \sqrt{1-b^2 u_O^2}$ and $\cos \phi_S = \sqrt{1-b^2 u_S^2}$.  For the light geodesic curvature, we use (\ref{27}) and (\ref{5}) to get 
\begin{align}
    k_g = -\frac{2Ma}{r^3} - \frac{2M^2a}{r^4} - \frac{3aM^3}{r^5} + \frac{8\alpha_0M^2a}{r^6} + \mathcal{O}\left(\frac{aM^4}{r^6}, \, \alpha_0^2\right). \label{35}
\end{align}
\begin{figure*}[!htb]
       \centering
    \begin{minipage}[t]{0.49 \textwidth}
        \centering
        \includegraphics[width = 1\textwidth]{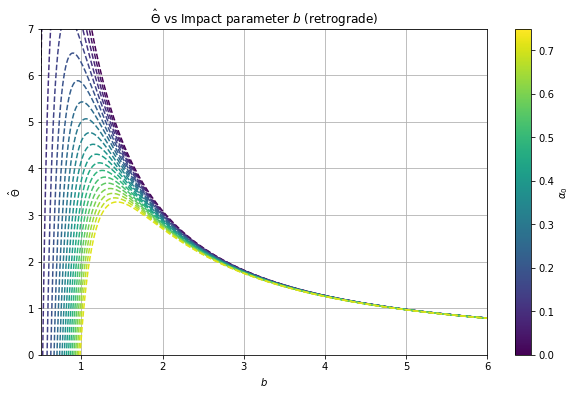}
    \end{minipage}
    \hfill
    \begin{minipage}[t]{0.49 \textwidth}
        \centering
        \includegraphics[width = 1\textwidth]{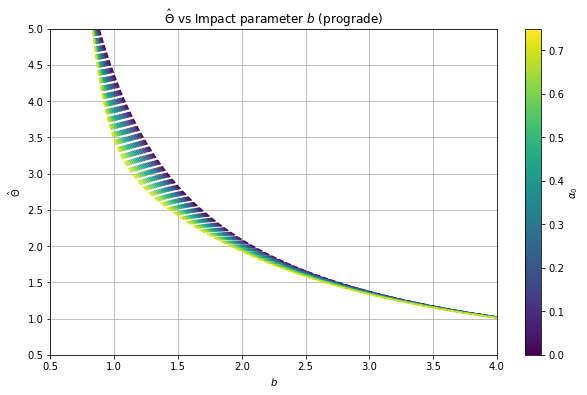}
    \end{minipage}
    \caption{\small{The weak deflection angle $\hat\Theta$ vs. impact parameter $b$ relation for retrograde (left) and prograde (right) motion of photons for the rotating regular black hole with Asymptotically Minkowski core, $x = 1, \ n =3$ for $0 \leq \alpha_0  \leq  0.73, \ a = 0.5$ and $M = 1$.}}
    \label{fig: 4}
\end{figure*}
\begin{figure*}[!htb]
       \centering
      \begin{minipage}[t]{0.49 \textwidth}
        \centering
        \includegraphics[width = 1\textwidth]{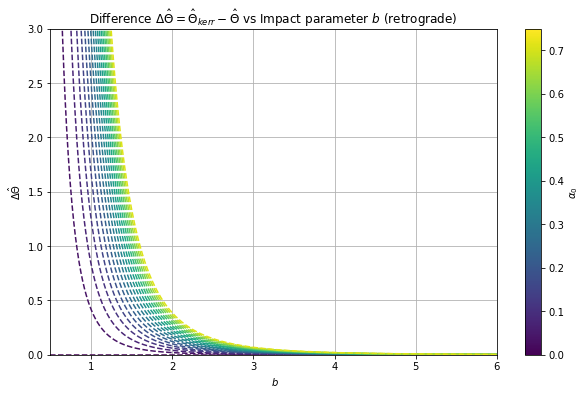}
    \end{minipage}
    \hfill
    \begin{minipage}[t]{0.49 \textwidth}
        \centering
        \includegraphics[width = 1\textwidth]{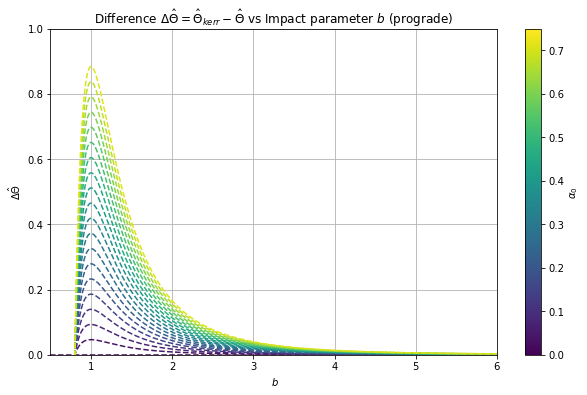}
    \end{minipage}
    \caption{\small{Deflection angle corrections $\delta{\hat\Theta}=\hat\Theta |_{Kerr} - \hat\Theta$ for the weak lensing around the rotating regular black hole with Asymptotically Minkowski core for retrograde (left) and prograde (right) motion of photons, $x = 1, \ n =3$ for $0\leq \alpha_0 \leq \ 0.73, \ a = 0.5$ and $M = 1$.}  }
    \label{fig: 5}
\end{figure*}
\hspace{3mm} Similarly, we solve for geodesic curvature integral using a photon orbit linear approximation as $r = b/\cos \vartheta$ and $l = b \tan \vartheta$ \cite{Ono:2017pie}.  Computing this integral, we obtain 
\begin{align}
\int_{S}^{O} k_g dl &= \int_{\phi_S}^{\phi_O} \left(-\frac{2Ma}{b^2}\cos\vartheta - \frac{2M^2 a}{b^3} \cos^2 \vartheta - \frac{3aM^3}{b^4} \cos^3 \vartheta + \frac{8 \alpha_0 M^2 a}{b^5} \cos^4 \vartheta \right) \notag \\ &= \left(\frac{3M^3 a}{b^4 }+ \frac{2aM}{b^2}\right)\left(\sqrt{1 - b^2 u_O^2}+ \sqrt{1-b^2 u_S^2}\right) \notag \\ &-\frac{M^3a}{b^4} \left[(1-b^2 u_O)^{3/2} + (1-b^2 u_S)^{3/2}\right] \notag \\ &+\left(\frac{aM^2 }{b^3} - \frac{3aM^2\alpha_O}{b^5}\right)\left[\arcsin\left(\sqrt{1-b^2 u_O^2}\right)+ \arcsin\left(\sqrt{1-b^2 u_S^2}\right)\right] \notag \\ &-\frac{aM^2\alpha_O}{4b^5}\left[u_O \sqrt{1-b^2 u_O^2} (1-2u_O^2b^2)+u_S \sqrt{1-b^2 u_S^2} (1-2u_S^2b^2)\right], \label{36}
\end{align}where we again used $\sin \phi_O = -\sqrt{1-b^2 u_O^2}$ and $\sin \phi_S = \sqrt{1-b^2 u_S^2}$.  Adding \eqref{34} and \eqref{36} and taking $u_O \rightarrow 0, \ u_S \rightarrow 0$, we get the deflection angle at the weak limit as 
\begin{align}
    \hat\Theta &\approx \frac{4M}{b} - \frac{15M^2 \alpha_0 \pi }{8b^4} \pm \left(\frac{4aM}{b^2} - \frac{3aM^2 \alpha_0 \pi}{b^5} \right). \label{37}
\end{align}
\hspace{3mm} In \cite{Hayward:2005gi}, Hayward proposed a regular black hole with an asymptotically de Sitter core where $m(r)$ in (\ref{5}) is 
\begin{align}
    m(r) = \frac{Mr^3}{r^3 + g^3}. \label{38}
\end{align}
\hspace*{3mm}The weak deflection angle by this type of regular black hole is derived in \cite{Jusufi:2018jof} as 
\begin{align}
    \hat\Theta = \frac{4M}{b} - \frac{15M\pi g^3}{8b^4} \pm \frac{4Ma}{b^2}. \label{39}
\end{align}
\hspace*{3mm} Setting $g^3 = x\alpha_0M^x = \alpha_0M$ in (\ref{4}), (\ref{37}) just becomes (\ref{39}).  This also shows the correspondence between this sort of regular black hole at $x = 1, \ n = 3$ with Hayward black hole in the weak field limit.\\
\hspace*{3mm}In Fig. \ref{fig: 4}, we see that the deviation parameter $\alpha_0$ has less effect for this sort of rotating regular black hole with $x=1, \ n=3$ than for $x = 2/3, \ n = 2$, which is particularly more evident in prograde motion.  For smaller values of $b$, we also see in Fig. \ref{fig: 5}  a similar trend in Figs. \ref{fig: 3}, but having lesser effect from the deviation parameter $\alpha_0$, than for the other sort of rotating regular blackhole with asymptotically Minkowski core discussed above.  Similar to Fig. \ref{fig: 3}, we observe that the deflection angle correction initially increases then decreases for different values of $\alpha_0$ for the prograde motion in Fig. \ref{5}.  

\section{Conclusion}
\hspace*{5mm} In this paper, we have investigated the deflection angle of light by rotating regular black holes with asymptotically Minkowski core as proposed by Ling and Wu \cite{Ling:2022vrv}. Using the Gauss-Bonnet Theorem as extended by Ono et al., the effects of the spin parameter $a$ and the parameter $\alpha_0$ were elucidated.  It revealed that the deflection angle of this sort of black hole is smaller than that of the Kerr black hole, but the difference vanishes over time as $b$ increases.  The trend also differs slightly between retrograde and prograde motion.  At this weak field limit, the deflection angle at certain values of $n$ and $\alpha_0$ is similar to Bardeen and Hayward black hole, which further supports the one-to-one correspondence of this sort of regular black hole and the regular black holes proposed by Bardeen and Hayward discussed by Ling and Wu.  

\newpage
\bibliographystyle{ieeetr}
\bibliography{references}

\end{document}